\documentclass[lettersize,journal]{IEEEtran}
\usepackage{ifpdf}
\usepackage{cite}
\ifCLASSINFOpdf
  \usepackage[pdftex]{graphicx}
\else
  \usepackage[dvips]{graphicx}
\fi
\usepackage{amsmath}
\usepackage{algorithm}
\usepackage{amsthm}
\usepackage{amsfonts}
\usepackage{amssymb}
\usepackage{color}
\usepackage{algorithmic}
\usepackage{array}
\ifCLASSOPTIONcompsoc
\usepackage[caption=false,font=normalsize,labelfont=sf,textfont=sf]{subfig}
\else
    \usepackage[caption=false,font=footnotesize]{subfig}
\fi
\usepackage{fixltx2e}
\usepackage{stfloats}
\ifCLASSOPTIONcaptionsoff
    \usepackage[nomarkers]{endfloat}
    \let\MYoriglatexcaption\caption
    \renewcommand{\caption}[2][\relax]{\MYoriglatexcaption[#2]{#2}}
\fi
\usepackage{url}
\usepackage{algorithmic}
\begin{document}

%\begin{verbatim}
%\newtheorem{definition}{Definition} 
%\newtheorem{theorem}{Theorem}
%\end{verbatim}

\title{Pareto-Aware Hierarchical Reinforcement Learning for Online Resource Allocation in RIS-assisted Large-Scale IoT Systems}

\author{Wenhan~Xu,~\IEEEmembership{Graduate Student Member,~IEEE,}
Jiashuo~Jiang,~\IEEEmembership{Member,~IEEE,}
%Cunhua~Pan,~\IEEEmembership{Senior Member,~IEEE,}
%Yulan~Yuan,~\IEEEmembership{Student Member,~IEEE,}
%Yuan~Wu,~\IEEEmembership{Senior Member,~IEEE,}
%Jiadong~Yu,~\IEEEmembership{Member,~IEEE,}
%Yuan~Wu,~\IEEEmembership{Senior Member,~IEEE,}
%Bo~Sun,~\IEEEmembership{Member,~IEEE,}
and~Danny~H.K.~Tsang,~\IEEEmembership{Life Fellow,~IEEE}% <-this % stops a space
%\thanks{This work was supported in part by Guangdong Provincial Project under Grant 2021JC02X149, in part by Guangzhou Municipal Science and Technology Project under Grant 2023A03J0011, in part by Guangzhou Municipal Key Laboratory on Future Networked Systems (024A03J0623), in part by Guangdong Provincial Key Laboratory of Integrated Communications, Sensing and Computation for Ubiquitous Internet of Things (No.2023B1212010007), in part by Research Grant of University of Macau under Grant MYRG-GRG2023-00083-IOTSC-UMDF, and in part by the Science and Technology Development Fund of Macau SAR under Grant FDCT 001/2024/SKL. \textit{(Corresponding author: Wenhan Xu.)} A preliminary version of this work appeared in the proceeding of IEEE 10th World Forum on Internet of Things (IEEE WF-IoT2024).}

\thanks{Wenhan Xu is with the Internet of Things Thrust, Hong Kong University of Science and Technology (Guangzhou), Guangzhou, Guangdong 511400, China (e-mail: wxube@connect.ust.hk).}% <-this % stops a space
\thanks{Jiashuo Jiang is with Department of Industrial Engineering and Decision Analytics, Hong Kong University of Science and Technology, Hong Kong 99907, China (e-mail: jsjiang@ust.hk).}% <-this % stops a space
\thanks{Danny H.K. Tsang is with the Internet of Things Thrust, Hong Kong University of Science and Technology (Guangzhou), Guangzhou, Guangdong 511400, China, and also with the Department of Electronic and Computer Engineering, The Hong Kong University of Science and Technology, Clear Water Bay, Hong Kong SAR, China (e-mail: eetsang@ust.hk).}
%\thanks{Cunhua Pan is with the National Mobile Communications Research Laboratory, Southeast University, Nanjing 210096, China.  (e-mail: cpan@seu.edu.cn).}% <-this % stops a space
%\thanks{Yulan Yuan is with the Internet of Things Thrust, Hong Kong University of Science and Technology (Guangzhou), Guangzhou, Guangdong 511400, China (e-mail: yyuan202@connect.hkust-gz.edu.cn).}
%\thanks{Yuan Wu is with The State Key Lab of Internet of Things for Smart City, and also with the Department of Computer and Information Science, University of Macau, Macao, China (e-mail: yuanwu@um.edu.mo).}% <-this % stops a space
%\thanks{Danny Hin-Kwok Tsang is with the Internet of Things Thrust, Hong Kong University of Science and Technology (Guangzhou), Guangzhou, Guangdong 511400, China (e-mail: eetsang@ust.hk).}
%\thanks{Digital Object Identifier 10.1109/JIOT.}
%\thanks{Copyright (c) 2023 IEEE. Personal use of this material is permitted. However, permission to use this material for any other purposes must be obtained from the IEEE by sending a request to pubs-permissions@ieee.org.}
}

\markboth{}%
{Shell \MakeLowercase{\textit{et al.}}: Bare Demo of IEEEtran.cls for IEEE Journals}



\maketitle

\begin{abstract}
With the rapid evolution of 5G and emerging 6G networks, reconfigurable intelligent surfaces (RIS) have become a critical technology for enhancing wireless communication scenarios. However, optimizing RIS-assisted multi-user systems typically introduces high-dimensional physical layer variables and non-convex Pareto-optimal rate sets, posing severe computational challenges for real-time applications. To address these limitations, this paper proposes a dimension-reduced, hierarchical reinforcement learning (RL) framework, termed Pareto-aware autoencoder-assisted RL (PAAERL), to optimize online resource allocation in RIS-assisted Internet of Things (IoT) networks. Our approach first substitutes high-dimensional continuous RIS beamforming variables with lower-dimensional weight vectors that strictly represent the Pareto-optimal frontier, theoretically avoiding geometric information loss across both convex and non-convex rate regions. To further mitigate the curse of dimensionality in dense networks, an autoencoder architecture is integrated to execute a secondary, data-driven compression phase, mapping the priority space into a highly condensed continuous latent action space. Extensive simulations conducted across practical communication scenarios, including multi-user mobile edge computing (MEC) networks, demonstrate that the proposed PAAERL framework drastically reduces offline training times, accelerates online policy convergence, and significantly decreases overall network costs compared to state-of-the-art benchmarks, underscoring its exceptional scalability and practical viability for next-generation intelligent IoT environments.

%With the rapid evolution of 5G and emerging 6G networks, Reconfigurable Intelligent Surfaces (RIS) have become a critical technology for enhancing wireless communication scenarios. However, optimizing RIS-assisted multi-user systems often involves high-dimensional variables and non-convex Pareto optimal sets. This paper proposes a low-scale hierarchical decision algorithm based on Reinforcement Learning (RL) to optimize online RIS-assisted IoT systems. By replacing high-dimensional RIS configuration variables with lower-dimensional weights representing the Pareto optimal set, the algorithm significantly reduces computational complexity while theoretically avoiding information loss in both convex and non-convex rate regions.

\end{abstract}


% \fi
\IEEEpeerreviewmaketitle

\section{Introduction}
Driven by the vision of next-generation B5G and emerging 6G wireless networks, the internet of things (IoT) is transitioning toward a paradigm of massive, ubiquitous connectivity and pervasive intelligence \cite{chien2025multi}. To fulfill the stringent quality-of-service (QoS) demands of modern application scenarios—such as ultra-reliable low-latency communications (URLLC) \cite{liu2023graph} and high-throughput edge processing—the integration of mobile edge computing (MEC) \cite{huang2024multi} and unmanned aerial vehicles (UAVs) \cite{huroon2025energy} has emerged as a promising architecture. While MEC servers alleviate the computational burden of resource-constrained IoT devices through task offloading, UAVs offer highly flexible, dynamic deployment to provide over-the-air coverage \cite{michailidis2024optimization}. However, the performance of these cooperative networks remains fundamentally constrained by severe path loss, unpredictable blockages, and multi-user interference inherent to high-frequency fading channels.

To mitigate these electromagnetic impediments, reconfigurable intelligent surfaces (RIS) have attracted profound attention as a revolutionary technology capable of transforming the wireless propagation environment from a passive, random medium into an active, software-programmable entity \cite{liu2021reconfigurable}. By dynamically adjusting the phase shifts and reflection amplitudes of a massive array of low-cost passive meta-elements, an RIS can intelligently reshape channel realizations and establish robust, effective line-of-sight (LoS) links between base stations (BSs) and mobile users (MUs) \cite{basar2024reconfigurable}. Despite these paradigm-shifting benefits, the joint optimization of active beamforming variables at the BS and passive reflection variables at the RIS introduces formidable mathematical hurdles. Traditional optimization frameworks—such as alternating optimization (AO) \cite{jiang2022interference} and successive convex approximation (SCA) \cite{wang2024double} are inherently iterative and scale poorly as the number of RIS elements and IoT devices expands. Solving these high-dimensional, tightly coupled non-convex problems within a single channel coherence interval inflicts prohibitive computational latency, rendering traditional model-driven paradigms impractical for real-time, online deployment in highly dynamic IoT environments.

To optimize multi-objective or multi-user networks, characterizing the Pareto-optimal boundary of the achievable rate region is fundamental. The most common approach utilized in existing literature is the linear scalarization technique, known as the weighted sum-rate maximization problem \cite{tang2025weighted}. By assigning fixed or dynamic priority weights to different users, the high-dimensional resource configuration is effectively mapped onto the Pareto frontier. Under the assumption of a convex rate region, adjusting the priority profile sequentially traces out the entire upper boundary of the feasible rate space via supporting hyperplanes. However, due to severe multi-user interference, transceiver hardware limitations, and non-linear cascade channel links, the true achievable rate region of a practical RIS-assisted multi-user network is inherently non-convex. Prior works \cite{jiang2021achievable,huang2020achievable} have pointed out that in non-convex rate spaces, linear weighted sum-rate maximization acts only as an outer approximation. Consequently, manipulating priority weights through traditional weighted sum-rate formulation inevitably experiences information loss, failing to capture or reach the optimal Pareto solution intervals located within the "dented" or non-convex portions of the boundary.

To bypass the computational overhead of online optimization, reinforcement learning (RL) has emerged as a powerful paradigm for real-time resource management \cite{peng2023energy}. RL agents learn near-optimal control policies through continuous interactions with the wireless environment, shift the heavy computational burden offline, and execute decisions rapidly online via neural network inference \cite{peng2025energy}. A twin-delayed deep deterministic policy gradient (TD3) framework was proposed in \cite{hashemi2022deep} to optimize the joint active/passive beamforming and finite blocklength allocation under realistic, non-ideal RIS amplitude impairments to address the stringent requirements of URLLC in industrial automation. Nonetheless, applying standard continuous RL algorithms such as deep deterministic policy gradient (DDPG), soft actor-critic (SAC), TD3 or proximal policy optimization (PPO) directly to RIS systems encounters severe action space explosion \cite{le2026learning}. A standard RIS configuration vector contains hundreds of continuous amplitude and phase control variables, leading to an incredibly vast exploration space. Under an uncompressed, high-dimensional action space, RL agents often suffer from slow convergence, gradient instability, or failing to discover meaningful reward paths entirely. While some existing works attempt to discretize the action space or use multi-agent RL (MARL) to split the variable workload \cite{zeng2025multiple}, they often break the coupled physical properties between active beamforming and passive reflection, yielding degraded system performance.
\subsection{Motivation and Contribution}
Differing from the aforementioned research, this paper aims to bridge the gap between model-driven geometric optimality and data-driven learning efficiency in the RIS scenario. Our previous work has made some contributions to similar optimization problems in RIS scenarios \cite{xu2023joint,xu2025reconfigurable,xu2024enhancing}. Rather than forcing a RL agent to learn the raw physical parameters of the RIS components directly which triggers a catastrophic action space explosion or relying on conventional weighted sum-rate profiles which suffer information loss in non-convex domains \cite{guo2020weighted}, we introduce a directional max-min scaling optimization problem. By constructing a bijective, direction-based mapping from a low-scale weight vector directly to the exact Pareto-optimal set, we collapse the search dimensionality without any theoretical loss of information. To further accelerate online execution in highly dynamic IoT environments, we seamlessly couple this geometric mapping with a data-driven autoencoder (AE) architecture \cite{erpek2021autoencoder}. We proposed a Pareto-aware autoencoder-aided RL (PAAERL) algorithm to solve the online resource allocation in RIS-assisted large scale IoT systems. The AE compresses the weight space into an ultra-low-dimensional latent representation, forcing the DRL agent to operate in a highly condensed action space while the accompanying decoder efficiently reconstructs the target priority vectors. This hybrid model-plus-data compression strategy enables a highly efficient hierarchical decision architecture tailored for real-time online optimization in complex RIS-assisted IoT systems.

%\subsection{Motivation and Contribution}

%Motivated by the literature review,

%The main contributions of our works are summarized as follows:
%\begin{itemize}
    %\item  
    %\item 
    %\item 
    %\item 
    %\item 
%\end{itemize}

\subsection{Notations}
In this paper, lowercase letters $x$, bold lowercase letters $\boldsymbol{x}$, and bold uppercase letters $\boldsymbol{X}$ denote the scalars, vectors, and matrices, respectively. $|x|$ is the absolute value of the scalar $x$. $x^*$ is the complex conjugate of the scalar $x$. The conditional expectation $\mathbb{E}\{x | y\}$ represents the expected value of the random variable $x$ given the condition $y$, computed with respect to the conditional distribution of $x$ given $y$. $\Vert \boldsymbol{x}\Vert$ is the 2-norm of vector $\boldsymbol{x}$. Furthermore, $\mathbf X^T$, $\mathbf X^H$, and $|\mathbf X|$ are the transpose, conjugate transpose, and determinant of a matrix $\mathbf X$, respectively. $diag(\boldsymbol{x})$ is a square matrix that has the vector $\boldsymbol{x}$ on the diagonal. The notation $\mathcal{O}(\cdot)$ denotes the asymptotic upper bound of an algorithm's computational complexity. Other primary notations in this paper are summarized in Table. \ref{Parameters_notation}.

\section{System model}

\begin{figure}[t]
\centering
\includegraphics[width=0.47\textwidth]{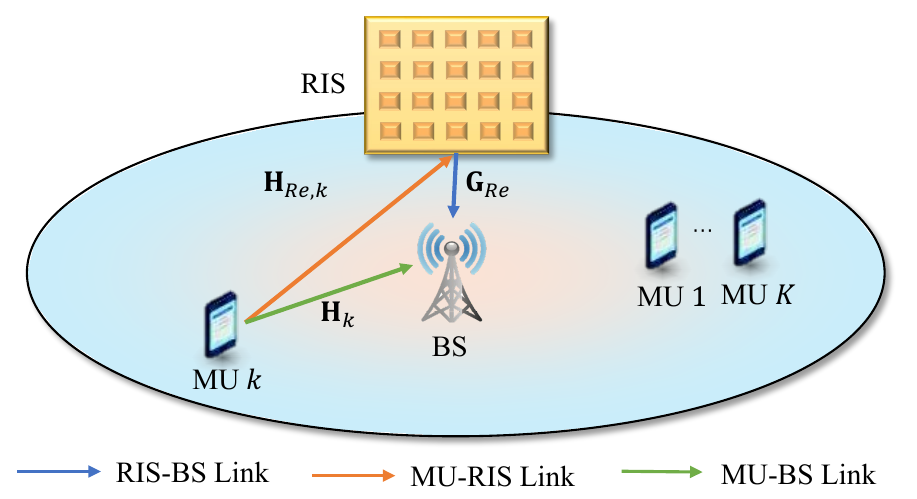}
\caption{The wireless communication model of an uplink RIS-aided IoT system.}
\label{sysmodel_HI}
\end{figure}

As illustrated in Fig. \ref{sysmodel_HI}, we consider an RIS-aided multi-user wireless communication system, where each MU is connected to a single BS. The deployment of RIS in this system model aims to enhance the quality of the channel of wireless communication by dynamically adjusting the amplitude and phase shift of the reflecting signal, thus improving resource utilization, reducing interference, and increasing the strength of the signal.

\begin{table}[t]
\renewcommand{\arraystretch}{1.3}
\caption{Parameters notation}
\label{Parameters_notation}
\centering
\begin{tabular}{c|c}
\hline
\bfseries Parameters & \bfseries Notation\\
\hline
$k,m$ & $k^{th}$ MU, and $m^{th}$ RIS element\\
$K,M$ & Number of MUs, and RIS elements\\
$t,\Delta T$ & Time-slot index and length of time slot\\
$\mathbf{f}_{k}$ & Transmit beamforming vector at the MU\\
${\mathbf H}_{k}$ & CSI from MU to BS\\
${\mathbf G}_{Re}$ & CSI from RIS to BS\\
${\mathbf H}_{Re,k}$ & CSI from MU to RIS\\
$\sigma^2$ & Variance of noise\\
$\mathbf{\Phi}$ & Diagonal RIS phase-shifting matrix\\
$\theta_{m},\beta_{m}$ & Phase and amplitude of RIS elements\\
$R_{k}$ & Achievable data rate (bit/s/Hz)\\
$\mathcal R$ & Achievable rate region\\
$\boldsymbol a, \boldsymbol s$ & Action and state of RL\\
$r$ & Reward of RL\\
\hline
\end{tabular}
\end{table}

\subsection{Channel Model}

We consider a system where a transmitted signal $x$ is reflected by an $M$-element RIS to a receiver $y$. The received signal is modeled as:

\begin{equation}
y = \left( \sum_{m=1}^{M} h_m g_m \beta_m e^{j\theta_m} \right)x + \eta
\end{equation}
where $h_m$ and $g_m$ are the first and second link channel coefficients, $\theta_m$ and $\beta_m$ represent the RIS phase and amplitude, respectively, and $\eta$ is the noise with variance $\sigma^2$.

As shown in Fig. \ref{sysmodel_HI}, we consider the uplink wireless communication model of MUs connected to a single BS separately and the intercellular interference is ignored. In the RIS-aided multi-user wireless communication system, each MU has $N_{MU}$ antenna, while the BS has $N_{BS}$ antennas. The controller of system can enhance both energy and spectral efficiency by designing the reflecting phase shifts of the RIS, which consists of $M$ reflection elements. $K$ MUs connected to the BS require computation offloading. The wireless signal ${\mathbf s}_{k}$ transmitted from the $k^{th}$ MU to the BS is given by
\begin{equation} 
	{\mathbf s}_{k} = {\mathbf f}_{k}{\mathbf{x}}_{k},
	\label{signal_transmitted_by_the_lth_BS}
\end{equation}
where $\mathbf f_{k}$ represents the transmit beamforming vector used by the $k^{th}$ MU for transmitting the symbol vector to the BS, and $\mathbf{x}_{k}$ represents the symbol transmitted from the $k^{th}$ MU to the BS. Hence, the received signal at the BS can be written as
\begin{equation}
	{{\mathbf y}} = \underbrace {\sum \limits _{k = 1}^{K} {{{\mathbf H}_{k}}{{\mathbf s}_{k}}} }_{{\mathrm{Direct~signal}}} +\underbrace {\sum \limits _{k = 1}^{K} { {\mathbf G}_{Re} \boldsymbol{\Phi} {\mathbf H}_{Re,k} } {{\mathbf s}_{k}}}_{{\mathrm{Reflecting~signal}}} + \boldsymbol{\eta}
\label{received_signal_vector_1},
\end{equation}
where $\mathbf{\Phi}=diag\{ \beta_{1} e^{j\theta_{1}},\beta_{2} e^{j\theta_{2}},...,\beta_{M} e^{j\theta_{M}}\}$ denotes the diagonal matrix of the RIS phase shifts and amplitudes, $\{\theta_{m},\beta_{m}\}$ is the phase shift and amplitude of the $m^{th}$ reflecting element of the RIS \cite{lu2021aerial}, and $\boldsymbol{\eta}$ is the noise vector with variance $\sigma^2$. As shown in Fig. \ref{sysmodel_HI}, the CSI for the MU-BS link, RIS-BS link, and MU-RIS link are ${\mathbf H}_{k}$, ${\mathbf G}_{Re}$, and ${\mathbf H}_{Re,k}$, respectively.
The MUs demonstrate quasi-static characteristics, while the perfect CSI varies with the time slot $t$. Each time slot has a fixed duration $\Delta T$, and all events occurring within the current time slot are processed in the subsequent time slot. Thus, the three CSI parameters ${\mathbf H}_{k}$, ${\mathbf G}_{Re}$, and ${\mathbf H}_{Re,k}$ here can be written as ${\mathbf H}_{k}(t)$, ${\mathbf G}_{Re}(t)$, and ${\mathbf H}_{Re,k}(t)$.
To simplify the expression, $\bar{\mathbf{H}}_{k}(t)$ is introduced:
\begin{equation}
	\bar{\mathbf{H}}_{k}(t)={\mathbf G}_{Re}(t) \boldsymbol{\Phi}(t) {\mathbf H}_{Re,k}(t)+\mathbf{H}_{k}(t).
 \label{g_mqk}
\end{equation}
In Eq. (\ref{received_signal_vector_1}), $\mathbf{y}(t)$ can be decomposed into received signal, extra interference, and additional noise when we focus on the transmitted signal from the $k^{th}$ MU to the BS. Therefore, the different components are shown as
\begin{equation}
\begin{aligned}
	\mathbf{y}(t) =
	& \sum \limits _{i = 1}^{K}{\bar {\mathbf H}_{i}(t){\mathbf f}_{i}(t){\mathbf x}_{i}(t)       }  +{\boldsymbol{\eta}}\\
	=& \underbrace {\bar {\mathbf H}_{k}(t){\mathbf f}_{k}(t){\mathbf x}_{k}(t) }_{{\mathrm {Signal}}}\\   
    +&\underbrace {\sum \limits _{i = 1,i \ne k}^{K}{\bar {\mathbf H}_{i}(t){\mathbf f}_{i}(t){\mathbf x}_{i}(t)       } }_{{\mathrm {Interference}}}   +{\boldsymbol{\eta}}, \forall k.
\end{aligned}
\end{equation}
Therefore, the uplink achievable data rate per unit of bandwidth from the $k^{th}$ MU to the BS is denoted by
\begin{equation}
R_{k} = \log\left(1 + \frac{||\bar{\mathbf H}_{k} \mathbf f_{k}||^2}{\sum^{K}_{i \ne k} ||\bar{\mathbf H}_{i} \mathbf f_{i}||^2 + \sigma^2}\right)
\label{r_ach}
\end{equation}

\subsection{Optimization Objective}

The primary objective is to minimize a cost function $f_{obj}$ of the achievable data rate $R_{k}$ and other optimization variables in the specific problem for MU $k$, defined by:

\begin{subequations}
    \begin{align}
		\mathcal P1:&\min_{\mathbb Z}\{f_{obj}(\boldsymbol R,\mathcal X)\} \nonumber\\
		\text {s.t.}\ &||{\mathbf f}_{k}(t)||^2 \leq P^{max}_k,\ \forall k, \\
        &\beta _{m}(t) \in [0,1],\ \forall m,\\
        &\theta _{m}(t) \in [0,2\pi),\ \forall m,\\
        &\text{other constraints},
    \end{align}
\end{subequations}
where $\mathcal X$ represents the other optimization variables in the specific problem,  $\mathbb Z=\{\mathbf f,\boldsymbol\beta,\boldsymbol\theta,\mathcal X\}$ represents all the optimization variables, $\mathbf f=\{{\mathbf f}_1,{\mathbf f}_2,...,{\mathbf f}_K\}$ represents all the beamforming vectors, $\boldsymbol{R} = [R_1,R_2,\dots, R_K]^T$ is the achievable rate vector, and $P^{max}_k$ is the maximum transmission energy for MU $k$.

\section{Proposed Hierarchical Decision Algorithm}
\label{Proposed_Algorithm}
\subsection{Dimensionality Reduction via Pareto Optimization}

In the considered RIS-assisted IoT uplink network, the system performance is characterized by the application-specific variables $\mathcal X$ (e.g., resource allocation or task offloading parameters) and the achievable rate vector $\boldsymbol{R} = [R_1,R_2,\dots, R_K]^T$, which is a function of the transmit beamforming vectors $\mathbf f=\{{\mathbf f}_1,{\mathbf f}_2,...,{\mathbf f}_K\}$, the RIS phase shift vector $\boldsymbol{\theta} = [\theta_1,\theta_2,\dots,\theta_M]^T$, and the RIS amplitude vector $\boldsymbol{\beta} = [\beta_1,\beta_2, \dots, \beta_M]^T$. The feasible set of all optimization variables $\mathbb Z=\{\mathbf f,\boldsymbol\beta,\boldsymbol\theta,\mathcal X\}$ is constrained by power budgets, RIS hardware limitations (where $0 \le \beta_m \le 1$ and $0 \le \theta_m < 2\pi$), QoS, and other specific requirements.

The achievable rate region of the system is defined as the set of all rate vectors that can be simultaneously attained:
\begin{equation}
	\mathcal{R} = \{ (R_1, \dots, R_K) \mid R_k \le \log_2 \left( 1 + \gamma_k(\mathbf f,\boldsymbol\beta,\boldsymbol\theta) \right),\forall k.\}
\end{equation}
where $\gamma_k$ denotes the signal-to-interference-plus-noise ratio (SINR) for user $k$:
\begin{equation}
    \gamma_k=\frac{||\bar{\mathbf H}_{k} \mathbf f_{k}||^2}{\sum^{K}_{i \ne k} ||\bar{\mathbf H}_{i} \mathbf f_{i}||^2 + \sigma^2}.
\end{equation}

Regardless of the non-convexity or complexity of the underlying objective function—whether it involves latency minimization, energy efficiency, or sum-rate maximization—any efficient operating point must lie on the Pareto boundary of $\mathcal{R}$. Formally, a rate vector $\boldsymbol{R}^*$ is Pareto optimal if there exists no other feasible $\boldsymbol{R} \in \mathcal{R}$ such that $R_k \ge R_k^*$ for all $k \in \{1, \dots, K\}$ and $R_j > R_j^*$ for at least one $j$.

The rationale for focusing on the Pareto boundary is twofold:
\begin{itemize}
\item Resource Efficiency: Any point in the interior of $\mathcal{R}$ is strictly sub-optimal, as at least one user's rate could be increased by reconfiguring $\{\mathbf f,\boldsymbol\beta,\boldsymbol\theta\}$ without detrimental effects on others.

\item Unified Representation: By characterizing the Pareto frontier, we decouple the physical layer resource interaction (governed by the RIS and beamforming) from the high-level policy or utility function.
\end{itemize}
Consequently, the high-dimensional optimization over $\{\mathbf f,\boldsymbol\beta,\boldsymbol\theta,\mathcal X\}$ can be equivalently reduced to finding a weight vector $\boldsymbol{\omega}$ or a priority profile that maps to a specific point on the Pareto boundary. This dimensionality reduction is vital for online IoT optimization, enabling real-time decision-making in environments characterized by complex variable coupling. Standard RIS optimization requires solving for beamforming vectors $\mathbf f_{k}$, RIS phases $\theta_{m}$, and amplitudes $\beta_{m}$. Instead of optimizing these directly, we propose using weight vectors $\omega$ that correspond to the Pareto-optimal solution set. The dimensionality of $\omega$ is far lower than the original RIS variables.

\subsection{Weighted Sum-rate Optimization}

While weighted sum-rate optimization is mature, it requires solving a traditional maximum weighted sum-rate problem at time slot $t$:
\begin{subequations}
    \begin{align}
		\mathcal P2:&\max_{{\mathbf f},\boldsymbol{\beta},\boldsymbol{\theta}}\sum_{k=1}^{K}\omega_{k}(t)R_{k}(t) \nonumber\\
		\text {s.t.}\ &||{\mathbf f}_{k}(t)||^2\leq P^{max}_k.\ \forall k,\\
        &\beta _{m}(t) \in [0,1],\ \forall m,\\
        &\theta _{m}(t) \in [0,2\pi),\ \forall m,
    \end{align}
\end{subequations}
where $\omega_{k}$ is the weight of $k^{th}$ MU at the BS. Clearly, given a set of $\{\omega_{k},\forall k\}$, a set of $\{{\mathbf f}_{k},\beta _{m},\theta _{m},\forall k,m\}$ can be solved. We can calculate $\{R_{k},\forall k\}$ using Eq. (\ref{r_ach}). By solving $\mathcal P2$, we obtained a set of mappings from $\{\omega_{k},\forall k\}$ to $\{R_{k},\forall k\}$.

For any given $\boldsymbol{\omega}$, the solution to this linear scalarization yields a rate vector $\boldsymbol{R}(\boldsymbol{\omega})$ that is theoretically guaranteed to lie on the Pareto boundary.

However, to justify the substitution of high-dimensional variables $\{\mathbf f,\boldsymbol\beta,\boldsymbol\theta,\mathcal X\}$ with the low-scale weight vector $\boldsymbol{\omega}$ in a hierarchical decision framework, it is imperative that the set of all points generated by $\boldsymbol{\omega}$ spans the entire Pareto frontier. We argue that when the achievable rate region $\mathcal{R}$ is a convex set, maximization of the weighted sum-rate is a sufficient characterization of the Pareto boundary.

Specifically, according to the Supporting Hyperplane Theorem, for any point $\boldsymbol{R}^*$ on the upper boundary of a convex set $\mathcal{R}$, there exists a supporting hyperplane with a non-negative normal vector $\boldsymbol{\omega}$ such that $\boldsymbol{\omega}^T \boldsymbol{R}^* \ge \boldsymbol{\omega}^T \boldsymbol{R}$ for all $\boldsymbol{R} \in \mathcal{R}$. This implies that every Pareto optimal point can be recovered by solving the weighted sum-rate problem with a corresponding weight profile. In this context, the weight vector $\boldsymbol{\omega}$ serves as a sufficient statistic for the RIS configuration, capturing all the degrees of freedom necessary to achieve any optimal resource allocation state.

While the interference-limited nature of RIS-assisted IoT systems often results in non-convex rate regions, the convex hull of $\mathcal{R}$ can be achieved through strategies such as time-sharing. Under such conditions, the mapping from $\boldsymbol{\omega}$ to $\mathcal{R}$ ensures complete coverage of the operationally relevant Pareto region, thereby validating the proposed dimensionality reduction.
\subsection{Generalization to Non-Convex Rate Regions and Directional Mapping}
While the convexity of the achievable rate region $\mathcal{R}$ can theoretically be ensured through time-sharing strategies, such approaches encounter significant practical impediments in RIS-assisted IoT systems. Specifically, the frequent reconfiguration of RIS phase shifts and amplitudes within a single transmission frame imposes a prohibitive overhead on the control link and may exceed the hardware switching rate limits. Consequently, within a quasi-static time slot, the rate region $\mathcal{R}$ is generally non-convex due to the complex coupling between the RIS reflection coefficients and the multi-user interference. In such non-convex scenarios, the conventional weighted sum-rate maximization may fail to capture the points residing in the "dented" or non-convex portions of the Pareto boundary.

To circumvent this limitation and ensure a comprehensive characterization of the Pareto set without relying on convexity assumptions, we propose an alternative optimization framework based on directional mapping. We observe that for any high-dimensional Pareto optimal point $\boldsymbol{R}^*$, the ray originating from the coordinate origin and passing through $\boldsymbol{R}^*$ is unique. This implies a one-to-one correspondence between the set of Pareto optimal points and the set of direction vectors in the positive orthant.

By leveraging this geometric property, we can parameterize the entire Pareto frontier using a normalized weight vector $\boldsymbol{\omega}$, which now represents the slope of the ray rather than the normal of a supporting hyperplane in the weighted sum-rate problem. To find the unique Pareto optimal point associated with a specific direction $\boldsymbol{\omega}$, we formulate the following optimization problem:

\begin{subequations}
    \begin{align}
		\mathcal P3:&\max_{{\mathbf f},\boldsymbol{\beta},\boldsymbol{\theta},R_s}R_s \nonumber\\
	\text {s.t.}\ & R_{k}(t)\geq\omega^{PT}_{k}R_s,\ \forall k,\label{R_pt}\\
         &||{\mathbf f}_{k}(t)||^2\leq P^{max}_k,\ \forall k,\\
        &\beta _{m}(t) \in [0,1],\ \forall m,\\
        &\theta _{m}(t) \in [0,2\pi),\ \forall m,
    \end{align}
\end{subequations}

In this formulation, $R_s$ acts as a scaling factor that pushes the rate vector along the direction $\{\omega^{PT}_{1},\omega^{PT}_{2},...,\omega^{PT}_{K}\}$ until it reaches the boundary of the feasible region $\mathcal{R}$. Unlike the weighted sum-rate method, this max-min scaling approach is guaranteed to reach any point on the Pareto boundary regardless of whether the region is convex or non-convex. This transformation effectively reduces the search space for the optimal RIS configuration from high-dimensional physical variables to a low-scale directional weight space, providing a robust theoretical foundation for our hierarchical RL decision algorithm.

\subsection{Problem Transformation and Solution Methodology}
Problem $\mathcal{P}3$ is a non-convex optimization problem due to the following reasons: First, the achievable rate $R_{k}$ is a non-convex function of the beamforming vectors $\mathbf f=\{{\mathbf f}_1,{\mathbf f}_2,...,{\mathbf f}_K\}$ and the RIS parameters $\{\boldsymbol{\beta}, \boldsymbol{\theta}\}$ due to the multi-user interference and the product of optimization variables in the effective channel. Second, the constant power constraint on $\mathbf f$ and the phase shift constraints on $\boldsymbol{\theta}$ define non-convex feasible sets.

To tackle these challenges, we observe that for fixed $\{\mathbf f, \boldsymbol{\beta}, \boldsymbol{\theta}\}$, the objective $R_s$ is linearly constrained, and for a fixed $R_s$, the problem reduces to a feasibility check. Exploiting this structure, we propose a two-layered solution framework.
\subsubsection{Outer Layer: Binary Search for $R_s$}  Since the achievable rate is bounded by the system's physical capacity, the optimal $R_s^*$ can be efficiently found using a binary search method over the interval $[0, R^{max}]$.

\subsubsection{Inner Layer: Feasibility Sub-problem} For a given candidate $R_s = \bar{R}_s$, we solve a feasibility problem to determine the existence of a configuration $\{\mathbf f, \boldsymbol{\beta}, \boldsymbol{\theta}\}$ that satisfies all constraints: 

\begin{subequations}
    \begin{align}
		\mathcal P4:&\max_{{\mathbf f},\boldsymbol{\beta},\boldsymbol{\theta}}0 \nonumber\\
	\text {s.t.}\ & R_{k}(t)\geq\omega^{PT}_{k}R_s,\ \forall k,\label{R_pt2}\\
         &||{\mathbf f}_{k}(t)||^2\leq P^{max}_k,\ \forall k,\\
        &\beta _{m}(t) \in [0,1],\ \forall m,\\
        &\theta _{m}(t) \in [0,2\pi),\ \forall m,
    \end{align}
\end{subequations}

Eq. (\ref{R_pt2}) can be simplified:
\begin{subequations}
    \begin{align}
	&R_{k}(t)\geq\omega^{PT}_{k}R_s,\ \forall k\\
	&\log\left(1 + \frac{||\bar{\mathbf H}_{k} \mathbf f_{k}||^2}{\sum^{K}_{i \ne k} ||\bar{\mathbf H}_{i} \mathbf f_{i}||^2 + \sigma^2}\right)\geq\omega^{PT}_{k}R_s,\ \forall k\\
         &\frac{||\bar{\mathbf H}_{k} \mathbf f_{k}||^2}{\sum^{K}_{i \ne k} ||\bar{\mathbf H}_{i} \mathbf f_{i}||^2 + \sigma^2}\geq2^{\omega^{PT}_{k}R_s}-1   ,\ \forall k,\\
        &||\bar{\mathbf H}_{k} \mathbf f_{k}||^2\geq (2^{\omega^{PT}_{k}R_s}-1)(\sum^{K}_{i \ne k} ||\bar{\mathbf H}_{i} \mathbf f_{i}||^2 + \sigma^2) ,\ \forall k.
    \end{align}
\end{subequations}
Therefore, the feasibility sub-problem can be simplified to:
\begin{subequations}
    \begin{align}
		\mathcal P4a:&\max_{{\mathbf f},\boldsymbol{\beta},\boldsymbol{\theta}} 0 \nonumber\\
	\text {s.t.}\ & ||\bar{\mathbf H}_{k} \mathbf f_{k}||^2\geq \Gamma_k(\sum^{K}_{i \ne k} ||\bar{\mathbf H}_{i} \mathbf f_{i}||^2 + \sigma^2) ,\ \forall k,\label{p4a_1}\\
         &||{\mathbf f}_{k}(t)||^2\leq P^{max}_k,\ \forall k,\\
        &\beta _{m}(t) \in [0,1],\ \forall m,\\
        &\theta _{m}(t) \in [0,2\pi),\ \forall m,
    \end{align}
\end{subequations}
where $\Gamma_k=2^{\omega^{PT}_{k}R_s}-1$ is a constant. The feasibility sub-problem $\mathcal{P}4a$ is inherently non-convex due to two primary factors: the strong coupling between the active beamforming vectors $\mathbf{f}$ and the passive RIS reflection coefficients (embedded within $\bar{\mathbf{H}}$), and the non-convexity of the lower-bounded quadratic terms in the Signal-to-Interference-plus-Noise Ratio (SINR) constraint. Firstly, we relax the feasibility problem of $\mathcal P4a$:

\begin{subequations}
    \begin{align}
		\mathcal P5:&\min_{{\mathbf f},\boldsymbol{\beta},\boldsymbol{\theta},\boldsymbol{d}} \sum_{k=1}^K d_k\nonumber\\
	\text {s.t.}\ & ||\bar{\mathbf H}_{k} \mathbf f_{k}||^2+d_k\geq \Gamma_k(\sum^{K}_{i \ne k} ||\bar{\mathbf H}_{i} \mathbf f_{i}||^2 + \sigma^2) ,\ \forall k,\\
         &||{\mathbf f}_{k}(t)||^2\leq P^{max}_k,\ \forall k,\\
        &\beta _{m}(t) \in [0,1],\ \forall m,\\
        &\theta _{m}(t) \in [0,2\pi),\ \forall m,
    \end{align}
\end{subequations}
where $d_k$ is the relax factor of the constraints Eq. (\ref{p4a_1}). To transform $\mathcal{P}5$ into a tractable form, we decouple the variables using an Alternating Optimization (AO) framework and convexify the constraints via Successive Convex Approximation (SCA). Furthermore, let $\mathbf{v} = [v_1, \dots, v_M]^T \in \mathbb{C}^{M \times 1}$ denote the RIS reflection vector, where $v_m = \beta_m e^{j\theta_m}$.

\subsubsection{Active Beamforming Optimization (Fixed RIS Coefficients)}
For a given RIS configuration $\mathbf{v}^{(n)}$ at the $n$-th iteration, the effective channel $\bar{\mathbf{H}}_k$ is fixed. The SINR constraint in $\mathcal{P}5$ can be expressed as:
\begin{equation}
||\bar{\mathbf{H}}_k \mathbf{f}_k||^2 +d_k \geq \Gamma_k \left( \sum_{i \neq k} ||\bar{\mathbf{H}}_i \mathbf{f}_i||^2 + \sigma^2 \right), \ \forall k.
\end{equation}

This constraint remains non-convex because the left-hand side is a convex quadratic function forming a lower bound. To address this, we apply the SCA technique by utilizing the first-order Taylor expansion to construct a global linear underestimator. For any given local point $\mathbf{f}_k^{(n)}$, the signal power is bounded by:
\begin{equation}
\begin{aligned}
||\bar{\mathbf{H}}_k \mathbf{f}_k||^2 &\geq 2\text{Re}\left\{ (\mathbf{f}_k^{(n)})^H \bar{\mathbf{H}}_k^H \bar{\mathbf{H}}_k \mathbf{f}_k \right\} - ||\bar{\mathbf{H}}_k \mathbf{f}_k^{(n)}||^2 \\
&\triangleq \mathcal{L}(\mathbf{f}_k, \mathbf{f}_k^{(n)}), \ \forall k.
\end{aligned}
\end{equation}

By replacing the non-convex term with its convex surrogate $\mathcal{L}(\mathbf{f}_k, \mathbf{f}_k^{(n)})$, and relaxing the strict equality power constraint to a convex inequality, which is known to be tight at optimality, the beamforming optimization sub-problem becomes:
\begin{equation}
\begin{aligned} 
\mathcal{P}5a: &\quad \min_{\mathbf{f}}  \sum_{k=1}^K d_k \\ 
\text{s.t.} \quad & \mathcal{L}(\mathbf{f}_k, \mathbf{f}_k^{(n)})+d_k \geq \Gamma_k \left( \sum_{i \neq k} ||\bar{\mathbf{H}}_i \mathbf{f}_i||^2 + \sigma^2 \right),\ \forall k, \\ 
& ||\mathbf{f}_k||^2 \leq P^{max}_k,\ \forall k. 
\end{aligned}
\end{equation}
Problem $\mathcal{P}5a$ is a standard convex quadratically constrained problem (QCP) that can be solved optimally using interior-point methods via solvers such as CVX.

\subsubsection{RIS Configuration Optimization (Fixed Beamforming)}

Next, we optimize the RIS reflection vector $\tilde{\mathbf{v}}$ while fixing the beamforming vectors $\{\mathbf{f}_k,\forall k\}$. The received signal power from user $k$ can be reformulated to isolate $\tilde{\mathbf{v}}$: 
\begin{equation}
\tilde{\mathbf{v}}=[v_1,...,v_M,v_{M+1}]^T=[\beta_{1} e^{j\theta_{1}},...,\beta_{M} e^{j\theta_{M}},1]^T.
\end{equation}
\begin{equation}
||\bar{\mathbf{H}}_k \mathbf{f}_k||^2=\sum_{i=1}^{M+1}\sum_{j=1}^{M+1}a_{k,i,j}v_i^*v_j,
\end{equation}%a is the complex conjugate of b.
where $v_i^*$ is the complex conjugate of $v_i$, $a_{k,i,j}$ is constant complex scalar, and $a_{k,i,j}=a_{k,j,i}^*$. The channel gain can be rewritten as:
\begin{equation}
||\bar{\mathbf{H}}_k \mathbf{f}_k||^2 = \tilde{\mathbf{v}}^H \mathbf{A}_{k} \tilde{\mathbf{v}},\ \forall k,
\end{equation}
where the corresponding semi-definite matrix:
\begin{equation}
\mathbf{A}_{k} = 
\begin{bmatrix} 
a_{k,1,1} & \cdots &a_{k,1,M+1} \\ 
\vdots & \ddots &\vdots \\
a_{k,M+1,1} & \cdots & a_{k,M+1,M+1}
\end{bmatrix},\ \forall k.
\end{equation}

Consequently, the SINR constraint with respect to $\tilde{\mathbf{v}}$ is formulated as:
\begin{equation}
\tilde{\mathbf{v}}^H \mathbf{A}_{k} \tilde{\mathbf{v}} \geq \Gamma_k \left( \sum_{i \neq k} \tilde{\mathbf{v}}^H \mathbf{A}_{i} \tilde{\mathbf{v}} + \sigma^2 \right)
\end{equation}

Similar to the beamforming sub-problem, the left-hand side introduces non-convexity. Applying SCA, we define the first-order lower bound around the previous iterate $\tilde{\mathbf{v}}^{(n)}$:
\begin{equation}
\begin{aligned}
\tilde{\mathbf{v}}^H \mathbf{A}_{k} \tilde{\mathbf{v}} &\geq 2\text{Re}\left\{ (\tilde{\mathbf{v}}^{(n)})^H \mathbf{A}_{k} \tilde{\mathbf{v}} \right\} - (\tilde{\mathbf{v}}^{(n)})^H \mathbf{A}_{k} \tilde{\mathbf{v}}^{(n)}\\
&\triangleq \mathcal{T}(\tilde{\mathbf{v}}, \tilde{\mathbf{v}}^{(n)},\mathbf{A}_{k}),\ \forall k.
\end{aligned}
\end{equation}

The physical limitations of the RIS elements require $|{v}_m| \leq 1$ for $m = 1, \dots, M$, and ${v}_{M+1} = 1$. Thus, the RIS sub-problem is given by:
\begin{equation}
\begin{aligned}
\mathcal{P}5b: &\min_{\tilde{\mathbf{v}}}  \sum_{k=1}^K d_k \\ 
\text{s.t.} & \mathcal{T}(\tilde{\mathbf{v}}, \tilde{\mathbf{v}}^{(n)},\mathbf{A}_{k})+d_k \geq \Gamma_k \left( \sum_{i \neq k} \tilde{\mathbf{v}}^H \mathbf{A}_{i} \tilde{\mathbf{v}} + \sigma^2 \right), \forall k, \\ 
& |{v}_m|^2 \leq 1, \quad m = 1, \dots, M, \\
& {v}_{M+1} = 1.
\end{aligned}
\end{equation}
This is a convex quadratically constrained quadratic problem (QCQP). The overall feasibility of $\mathcal{P}4$ is determined by alternately solving $\mathcal{P}5a$ and $\mathcal{P}5b$. Since the SCA lower bounds guarantee that the objective value is non-decreasing and bounded, the AO process is guaranteed to converge to a stationary point of the feasibility problem.

By iteratively updating $R_s$ and solving the corresponding feasibility sub-problems, the algorithm converges to the Pareto optimal boundary point along the ray $\boldsymbol{\omega}^{PT}$. This approach ensures that the high-dimensional resource allocation is uniquely and optimally mapped to the low-scale weight space, facilitating real-time execution in dynamic IoT environments. This approach ensures no theoretical information loss regardless of region convexity.

Based on the preceding analysis, we have established a deterministic mapping $\mathcal{M}: \boldsymbol{\omega}^{PT} \to \boldsymbol{R}^*$, where $\boldsymbol{R}^*$ denotes the unique Pareto optimal rate vector obtained by solving problem $\mathcal{P}3$ for a given direction $\boldsymbol{\omega}^{PT}$. Since the directional mapping $\mathcal{M}$ is surjective over the entire Pareto frontier of the achievable rate region $\mathcal{R}$, it ensures that every efficient operating point of the RIS-assisted IoT system can be uniquely represented by a low-dimensional weight vector $\boldsymbol{\omega}^{PT}$.

This geometric correspondence facilitates a significant paradigm shift in the system's optimization strategy. Specifically, we can substitute the high-dimensional and tightly coupled physical layer variables, including the beamforming vectors $\mathbf f$, the RIS phase shifts $\boldsymbol{\theta}$, and the reflection amplitudes $\boldsymbol{\beta}$, with the low-scale weight vector $\boldsymbol{\omega}^{PT}$ as the primary decision variable. 

Crucially, this dimensionality reduction is information-lossless from an optimization perspective. Given that any physically meaningful objective function (e.g., minimizing system latency, maximizing energy efficiency, or total sum-rate) is typically a non-decreasing function of the individual user rates $R_{k}$, the global optimum for such objectives is mathematically guaranteed to reside on the Pareto boundary. Consequently, by searching within the low-dimensional space of $\boldsymbol{\omega}^{PT}$ instead of the original high-dimensional variable space, we can attain the global optimum while drastically reducing the search space complexity. This transformation provides the theoretical foundation for our hierarchical decision framework, where a high-level RL agent can focus on optimizing the weight profile $\boldsymbol{\omega}^{PT}$, while the low-level physical configurations are efficiently handled by the directional mapping $\mathcal{M}$. 

In the RL framework, the action space is $\boldsymbol a=\{\boldsymbol\omega^{PT},\mathcal X\}$, where $\mathcal X$ represents the other optimization variables in the specific problem. The state space consists of channel state information ${\mathbf H}_{k}, {\mathbf G}_{Re}, {\mathbf H}_{Re,k}$ and other state information $\mathcal Y$ in the specific problem:
\begin{equation}
\boldsymbol s=\{{\mathbf H}_{k}, {\mathbf G}_{Re}, {\mathbf H}_{Re,k},\mathcal Y,\forall k\}.
\end{equation}
The reward function as the negative of the objective in $\mathcal P1$ at time slot $t$: $r=-f_{obj}(\boldsymbol R,\mathcal X)$. Therefore, the Pareto-aware hierarchical decision algorithm using RL (PARL) is shown as Algorithm \ref{PARL}. 
\begin{algorithm}[t]
	\caption{Pareto-Aware Hierarchical Decision Algorithm using RL with PPO agent for Online RIS-aided IoT system}
        \renewcommand{\algorithmicrequire}{\textbf{Result:}}
	\renewcommand{\algorithmicensure}{\textbf{Input:}}
	\begin{algorithmic}[1]
        \ENSURE Upper limit of episodes, upper limit of steps per episode, learning factor.
		\REQUIRE The policy of RL: $PPO(\boldsymbol s,\boldsymbol a)$.
        \STATE Action space: $\boldsymbol a=\{\boldsymbol\omega^{PT},\mathcal X\}$.
		\STATE State space: $\boldsymbol s=\{{\mathbf H}_{k}, {\mathbf G}_{Re}, {\mathbf H}_{Re,k},\mathcal Y,\forall k\}$.
		\STATE Initialize a random policy $PPO(\boldsymbol s,\boldsymbol a)$ for each $\boldsymbol s$ and $\boldsymbol a$.
		\REPEAT 
            \STATE Initialize $\boldsymbol a^{(0)}$ with a random feasible action.
            \STATE Initialize $\boldsymbol s^{(0)}$ with $\boldsymbol a^{(0)}$ and system settings.
            \STATE $t=0$.
			\REPEAT
                    \STATE Select the next action $\boldsymbol a^{(t+1)}$ from the current policy $PPO$ and current state $\boldsymbol s^{(t)}$ using PPO method.
                    \STATE Perform the next action $\boldsymbol a^{(t+1)}$ by solving $\mathcal P3$:
                    \STATE Set $R^{min}_s=0$, and find $R^{max}_s$ when $\mathcal P4$ has no feasible solution.
                    \REPEAT
                        \STATE $R_s=\frac{R^{min}_s+R^{max}_s}{2}$
                        \IF{$\mathcal P4$ has feasible solution} 
                            \STATE $R^{min}_s=R_s$
                        \ELSE 
                            \STATE $R^{max}_s=R_s$
                        \ENDIF
                    \UNTIL $R^{max}_s-R^{min}_s<\epsilon$, $\epsilon$ is the accuracy requirement.
                    \STATE Update the next state $\boldsymbol s^{(t+1)}$ with $\boldsymbol a^{(t+1)}$.
                    \STATE Observe the reward $r^{(t+1)}$ from $\{\boldsymbol a^{(t+1)},\boldsymbol s^{(t+1)}\}$.
                    \STATE Store $\{\boldsymbol a^{(t+1)},\boldsymbol s^{(t+1)},r^{(t+1)}\}$ into the trajectory buffer $\mathcal{G}$. Update RL policy network via PPO using collected trajectory batch $\mathcal{G}$.
                    \STATE $t=t+1$.
			\UNTIL Upper limit of steps per episode.
            \STATE Clear the trajectory buffer $\mathcal{G}$.
		\UNTIL Upper limit of episodes or convergence.
	\end{algorithmic}
	\label{PARL}
\end{algorithm}

\subsection{Secondary Compression of Action Space using Autoencoder}

\begin{figure}[t]
\centering
\includegraphics[width=0.47\textwidth]{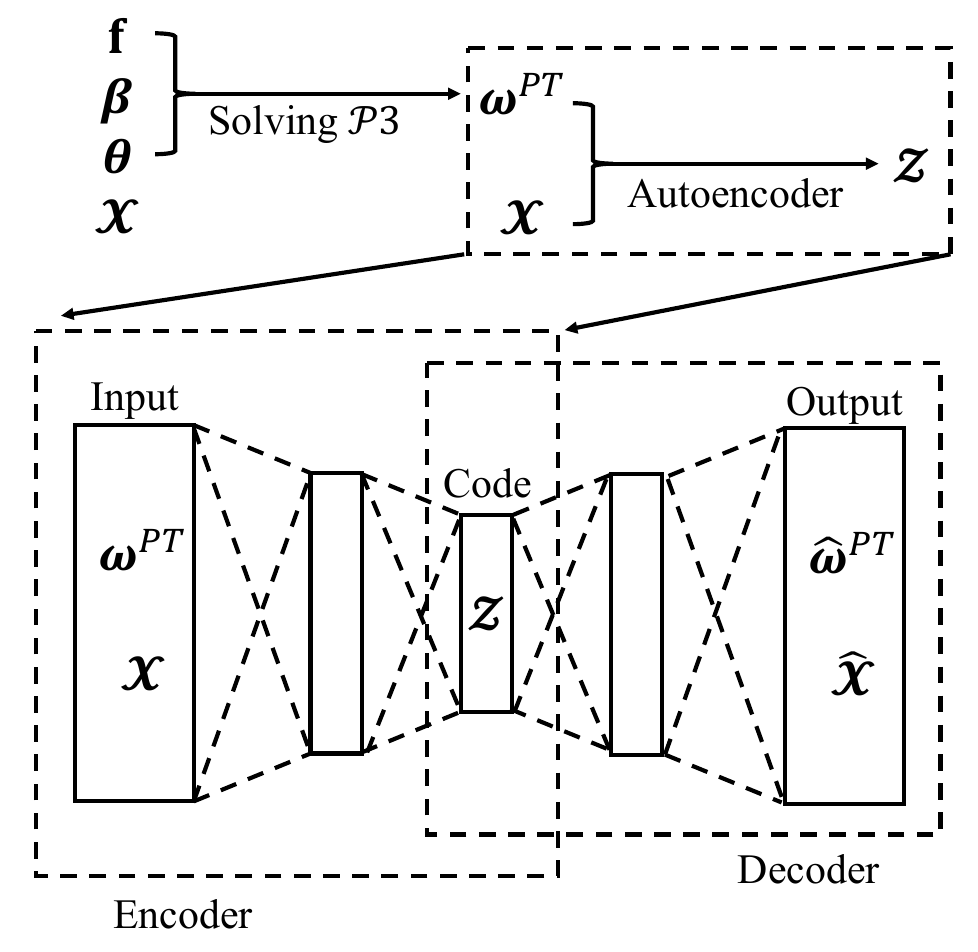}
\caption{The structure of autoencoder.}
\label{AE}
\end{figure}

While the direction-based priority mapping $\mathcal{M}: \boldsymbol{\omega}^{PT} \to \boldsymbol{R}^*$ proposed in the previous subsection effectively collapses the high-dimensional physical space of active beamforming and passive reflection coefficients into a $K$-dimensional priority vector, directly deploying RL over $\boldsymbol{\omega}^{PT}$ may still pose challenges in massive IoT networks where $K$ is large. Furthermore, spatial correlation among user distributions and recurring environmental dynamics often introduce statistical redundancies within the priority vector space. To exploit these latent statistical regularities and achieve an ultra-low-dimensional action representation for the RL agent, we introduce a data-driven secondary compression framework utilizing an autoencoder (AE) as shown in Fig. \ref{AE}.

Unlike model-driven geometric projections, the AE architecture is entirely data-driven and relies on extracting low-dimensional continuous manifolds from optimal configuration profiles. Consequently, the secondary compression framework necessitates a dedicated offline pre-training phase.

We first execute the primary directional mapping algorithm over a diverse set of random channel realizations and varying network states. By iteratively performing the binary search over $R_s$ and solving the convex feasibility sub-problems $\mathcal{P}5a$ and $\mathcal{P}5b$, we accumulate a comprehensive offline dataset $\mathcal{D}_{train} = \{\{\boldsymbol\omega^{PT},\mathcal X\}^{(1)}, \{\boldsymbol\omega^{PT},\mathcal X\}^{(2)}, \dots, \{\boldsymbol\omega^{PT},\mathcal X\}^{(N_{sa})}\}$. Each entry represents an optimized $K+K_{\mathcal X}$-dimensional target priority profile residing strictly on the Pareto boundary, where $K_{\mathcal X}$ is the dimension of $\mathcal X$

The pre-trained dataset is subsequently used to optimize a symmetric deep neural network structure comprising an encoder and a decoder.

The encoder ($\mathcal{E}_{\phi}$): Takes the optimized, uncompressed priority vector $\mathbf{X} = \{\boldsymbol\omega^{PT},\mathcal X\} \in \mathbb{R}^{K+K_{\mathcal X}}$ as input. Through a series of fully connected layers parameterized by weights and biases $\phi$, it projects the vector into an ultra-low-dimensional bottleneck layer, yielding the latent continuous code vector $\mathcal{Z} \in \mathbb{R}^{d \times 1}$, where $d < K+K_{\mathcal X}$. This latent vector $\mathcal{Z}$ acts as the condensed, highly efficient action space explored directly by the high-level RL policy network.

The decoder ($\mathcal{D}_{\psi}$): Tasked with reversing the compression operation to maintain physical consistency. Parameterized by weights $\psi$, the decoder takes the latent representation $\mathcal{Z}$ from the bottleneck layer and maps it back to the original dimensionality, generating a reconstructed priority vector $\hat{\mathbf{X}} =  \{\hat{\boldsymbol\omega}^{PT},\hat{\mathcal X}\} \in \mathbb{R}^{K+K_{\mathcal X}}$.

The network parameters $\{\phi, \psi\}$ are jointly optimized offline by minimizing the mean squared error (MSE) reconstruction loss via backpropagation:

\begin{equation}
\min_{\phi, \psi} \quad \frac{1}{N_{sa}} \sum_{n=1}^{N_{sa}} \|\mathbf{X}^{(n)} - \mathcal{D}_{\psi}(\mathcal{E}_{\phi}(\mathbf{X}^{(n)}))\|^2,
\end{equation}
where $N_{sa}$ represents the size of the dataset. Once the offline training converges, the frozen Decoder network is embedded directly into the online execution loop as an intermediate translation tier between the RL policy and the physical layer optimization module.

The precise cascading mechanics of the online hybrid decision pipeline operate as follows:

\begin{itemize}
    \item  Macro-Decision Generation: At each time slot $t$, the DRL agent observes the low-dimensional network state space and outputs a continuous latent action vector $\mathcal{Z}(t)$ from its squeezed action space.
    \item Data-Driven Decompression: The latent code $\mathcal{Z}(t)$ is instantly fed into the feedforward Decoder network $\mathcal{D}_{\psi}$, which outputs the reconstructed priority profile $\{\hat{\boldsymbol{\omega}}^{PT}(t),\hat{\mathcal X}(t)\}$.
    \item Model-Driven Projection: To insulate the system from neural reconstruction noise, $\{\hat{\boldsymbol{\omega}}^{PT}(t),\hat{\mathcal X}(t)\}$ is passed directly to the model-driven feasibility loop. The system solves the inner problems $\mathcal{P}4$ via the algorithm mentioned above.
\end{itemize}

By sequentially chaining the statistical dimensionality reduction of the autoencoder with the exact geometric mapping of the directional max-min problem, the framework creates a robust "firewall" against the curse of dimensionality. The RL agent benefits from fast training convergence in an optimized continuous action space ($\mathcal{Z}$), while the physical layer constraints are guaranteed to satisfy strict Pareto optimality conditions without information loss if AE can perfectly replace the original parameters with a small number of parameters. If AE cannot perfectly replace parameters, it will only cause a small amount of performance loss.

\begin{algorithm}[t]
	\caption{Pareto-Aware Autoencoder-aided RL (PAAERL) Algorithm}
        \renewcommand{\algorithmicrequire}{\textbf{Result:}}
	\renewcommand{\algorithmicensure}{\textbf{Input:}}
	\begin{algorithmic}[1]
        \ENSURE Upper limit of episodes, upper limit of steps per episode, learning factor.
		\REQUIRE The policy of RL: $PPO(\boldsymbol s,\boldsymbol a)$.
        \STATE Action space: $\boldsymbol a=\{\mathcal{Z}\}$.
		\STATE State space: $\boldsymbol s=\{{\mathbf H}_{k}, {\mathbf G}_{Re}, {\mathbf H}_{Re,k},\mathcal Y,\forall k\}$.
		\STATE Initialize a random policy $PPO(\boldsymbol s,\boldsymbol a)$ for each $\boldsymbol s$ and $\boldsymbol a$.
		\REPEAT 
            \STATE Initialize $\boldsymbol a^{(0)}$ with a random feasible action.
            \STATE Initialize $\boldsymbol s^{(0)}$ with $\boldsymbol a^{(0)}$ and system settings.
            \STATE $t=0$.
			\REPEAT
                    \STATE Select the next action $\boldsymbol a^{(t+1)}$ from the current policy $PPO$ and current state $\boldsymbol s^{(t)}$ using PPO method.
                    \STATE Use AE to obtain the original parameters $\{\boldsymbol\omega^{PT},\mathcal X\}^{(t+1)}$ from $\mathcal{Z}^{(t+1)}$.
                    \STATE Perform the next action $\boldsymbol a^{(t+1)}$ by solving $\mathcal P3$.
                    \STATE Update the next state $\boldsymbol s^{(t+1)}$ with $\boldsymbol a^{(t+1)}$.
                    \STATE Observe the reward $r^{(t+1)}$ from $\{\boldsymbol a^{(t+1)},\boldsymbol s^{(t+1)}\}$.
                    \STATE Store $\{\boldsymbol a^{(t+1)},\boldsymbol s^{(t+1)},r^{(t+1)}\}$ into the trajectory buffer $\mathcal{G}$. Update RL policy network via PPO using collected trajectory batch $\mathcal{G}$.
                    \STATE $t=t+1$.
			\UNTIL Upper limit of steps per episode.
            \STATE Clear the trajectory buffer $\mathcal{G}$.
		\UNTIL Upper limit of episodes or convergence.
	\end{algorithmic}
	\label{PAAERL}
\end{algorithm}

\subsection{Complexity}
The offline training complexity of the PAAERL algorithm is primarily governed by the synchronization between data-driven neural network backpropagation and the model-driven geometric projection loops. The total asymptotic training complexity of the PAAERL algorithm can be expressed as:
\begin{equation}
\mathcal{O}(\text{PAAERL}) = \mathcal{O}\left( N_I N_B \left( N_{RL} + N_{PA} \right) \right)
\end{equation}
where $N_I$ denotes the total number of training iterations, $N_B$ represents the number of gradient updates executed per iteration, $N_{RL}$ is the floating-point operations required for a single training step, and $\mathcal{O}(N_{PA})$ is the complexity of binary search and solve $\mathcal P4$. The complexity is given by:
\begin{equation}
\mathcal{O}(N_{PA}) = \mathcal{O}\left( \log_2\left(\frac{\epsilon_0}{\epsilon}\right) \cdot N_{\mathcal{P}4} \right)
\end{equation}
where $\log_2(\epsilon_0 / \epsilon)$ represents the maximum number of bisection steps required for the binary search to achieve an accuracy tolerance of $\epsilon$, and $\mathcal{O}(N_{\mathcal{P}4})$ denotes the complexity required to solve sub-problems $\mathcal{P}4$ for $K$ MUs.

In most RL setups, the number of required iterations $N_I$ typically scales polynomially with the uncompressed state-action product $N_{\text{state}}N_{\text{action}}$. However, by cascading the data-driven autoencoder's structural condensation with the exact boundary projection of problem $\mathcal{P}3$, PAAERL dramatically truncates the required exploration volume.

\section{Simulation and Application}

The theoretical framework established in the previous sections possesses a system-agnostic quality, rendering it applicable to a broad spectrum of multi-user communication scenarios incorporating RIS. In principle, any optimization objective that is a monotonic function of the users' achievable rates can be efficiently solved within this low-scale hierarchical structure. To demonstrate the versatility and extensibility of the proposed method, we consider its application to a RIS-assisted MEC system as a representative case study.

In a typical RIS-assisted MEC environment, multiple MUs generate computation-intensive tasks that can be either processed locally or offloaded to an edge server via RIS-augmented uplink channels. The optimization problem in such a system is significantly more complex, as it involves the joint allocation of communication resources (beamforming vectors $\mathbf f$ and RIS coefficients $\{\boldsymbol{\theta}, \boldsymbol{\beta}\}$) and secondary computing resources (local CPU frequencies, and edge server scheduling). 

In dynamic operational scenarios, new computation tasks arrive stochastically at each time slot according to a task arrival process. Accumulated tasks at each MU form a local buffer queue. Each MU is assigned an achievable uplink transmission rate determined by the joint beamforming parameters $\mathbf{f}$, $\boldsymbol{\theta}$, and $\boldsymbol{\beta}$. MUs transmit data from the task buffer queue to the edge server at their allocated uplink rates while concurrently executing local computation at a separate CPU processing rate. The primary objective is to minimize the accumulated weighted cost combining the spatial average latency $\bar{D}(t)$ and energy consumption $\bar{E}(t)$ at time slot $t$. The spatial average latency across $K$ active MUs is expressed as:
\begin{equation}
\bar{D}(t) = \frac{1}{K} \sum_{k=1}^K D_k(t),
\end{equation}
where $D_k(t)$ denotes the task execution latency experienced by the $k^{th}$ MU at time slot $t$. Concurrently, $\bar{E}(t)$ captures the aggregate energy overhead, encompassing local CPU computation energy at each MU, and edge computation energy consumed by the server. The joint dynamic optimization problem is formulated as follows:
\begin{subequations}
    \begin{align}
		\mathcal P6:&\min_{\mathbf f,\boldsymbol\beta,\boldsymbol\theta,\boldsymbol c^{MU},\boldsymbol c}\sum_{t=1}^T\lambda^{t-1}\left(\xi_1 \bar{D}(t) + \xi_2 \bar{E}(t)\right) \nonumber\\
		\text {s.t.}\ &||{\mathbf f}_{k}(t)||^2 \leq P^{max}_k,\ \forall k,t\\
        &\beta _{m}(t) \in [0,1],\ \forall m,t\\
        &\theta _{m}(t) \in [0,2\pi),\ \forall m,t\\
        &c_k^{MU}(t)\leq c_k^{MU,max},\ \forall k,t\\
        &c(t)\leq c^{max}, \forall t,
    \end{align}
\end{subequations}
where $T$ is the total number of time slots, $\lambda \in [0, 1)$ is the temporal discount factor, and $\xi_1, \xi_2 \ge 0$ are design weighting parameters prioritizing delay and energy metrics, respectively. Furthermore, $c_k^{MU}(t)$ denotes the local computation processing rate (bit/s) at the $k^{th}$ MU subjected to the local computing capacity constraint $c_k^{MU,max}$, and $c(t)$ represents the overall processing rate assigned by the edge server, subject to its peak computing capacity constraint $c^{max}$. Since the problem operates over a predetermined finite time horizon, choosing a discount factor $\lambda$ close to $1$ ensures long-term optimization stability while maintaining convergence.

To solve problem $\mathcal{P}6$ via RL, we align the policy objective by defining the immediate reward function at time slot $t$ as:
\begin{equation}
r^{(t)} = -\xi_1 \bar{D}(t) - \xi_2 \bar{E}(t).
\end{equation}
Maximizing the cumulative expected reward $\mathbb{E}\left[\sum_{t=1}^T \lambda^{t-1} r^{(t)}\right]$ corresponds directly to minimizing the original objective function in $\mathcal{P}6$. Detailed queueing models and MEC system formulations built on similar dynamic principles can be found in relevant research works \cite{jia2022lyapunov,bi2021lyapunov,di2021dynamic}.

%The objective function of problem $\mathcal P6$ represents an accumulated weighted cost function of latency energy. Therefore, the reward functions for each time slot is $r = -\xi_1 \bar{D}(t) - \xi_2 \bar{E}(t)$, which ensures that the RL maximizes the cumulative reward, which corresponds exactly to the negative of the original objective function. Detailed MEC system models already exist in many relevant research works\cite{jia2022lyapunov,bi2021lyapunov,di2021dynamic}.

By employing the proposed hierarchical decision algorithm, the high-level RL agent is tasked with navigating the low-dimensional space of rate proportions $\boldsymbol{\omega}^{PT}$ and high-level offloading strategies, rather than dealing with the microscopic adjustments of RIS elements. Since the latency and energy consumption are directly governed by the achievable uplink rate $R_{k}$, the optimal resource configuration for any given offloading decision must still reside on the Pareto boundary of the rate region. 

Consequently, the RL agent can effectively delegate the physical layer coordination to the $\mathcal{P}3$ mapping, ensuring that for any selected strategy, the RIS and beamforming parameters are always in their Pareto-optimal state. This separation of concerns not only stabilizes the learning process of the RL agent in dynamic IoT environments but also ensures that the system achieves the theoretical lower bound of the cost function with significantly reduced computational overhead. Such a framework can be further extended to other emerging architectures, including STAR-RIS and UAV-relayed networks, where high-dimensional coupling is a persistent bottleneck.

To validate the efficacy, convergence properties, and resource-allocation performance of the proposed Pareto-aware autoencoder-aided RL (PAAERL) framework, we establish a comprehensive simulation environment for a multi-user RIS-assisted network. For a fair and rigorous performance evaluation, we compare PAAERL against five baseline paradigms. These benchmark algorithms are strategically categorized based on their optimization criteria (weighted sum-rate vs. Pareto-optimality) and their dimensionality reduction strategies (model-driven, data-driven, or uncompressed), as detailed below:

\begin{itemize}
    \item RL: This baseline represents the standard model-free, end-to-end data-driven approach widely utilized in existing literature. In this scheme, the RL agent directly observes the raw channel state information and outputs the high-dimensional physical layer configuration parameters without any prior space compression.
    \item WSRRL: The weighted sum-rate assisted RL algorithm utilizes a conventional model-driven compression method. Instead of exploring the physical variable space directly, the DRL agent outputs a priority weight vector $\boldsymbol{\omega}$. A low-level optimization solver then maximizes the conventional weighted sum-rate objective function based on these weights to determine $\mathbf f$, $\boldsymbol{\theta}$, and $\boldsymbol{\beta}$.
    \item PARL: The Pareto-aware RL algorithm represents an ablation version of our proposed framework that isolates the model-driven tier. It exclusively employs the directional max-min scaling optimization problem $\mathcal{P}3$ developed in Section \ref{Proposed_Algorithm} to map the $K$-dimensional weight vector $\boldsymbol{\omega}^{PT}$ directly onto the true Pareto-optimal frontier.
    \item AERL: The autoencoder-assisted RL baseline utilizes a purely data-driven action-embedding paradigm. It does not employ any communication-theoretic or geometric optimization models. Instead, an offline pre-trained Autoencoder compresses the raw, high-dimensional physical variables ($\mathbf{f}, \boldsymbol{\beta}, \boldsymbol{\theta}$) directly into a compact continuous latent code $\mathcal{Z}$. The DRL agent executes its exploration and policy updates entirely within this unconstrained latent space, and the decoder maps $\mathcal{Z}$ back into the physical layer parameters. This benchmark serves to contrast a purely black-box AI dimensionality reduction approach against our hybrid model-plus-data compression pipeline.
    \item WSRAERL: The weighted sum-rate autoencoder-assisted RL algorithm is a hybrid benchmark that combines conventional linear scalarization with neural network compression. In this scheme, the framework first maps the resource configuration space to a $K$-dimensional priority weight vector $\boldsymbol{\omega}$, which is subsequently compressed into a lower-dimensional latent representation using a data-driven Autoencoder. The DRL agent outputs a condensed action that is decoded into $\boldsymbol{\omega}$ and evaluated via the conventional weighted sum-rate optimization solver.
\end{itemize}
\subsection{Performance comparison}
In this section, we evaluate the performance of the proposed PAAERL framework against the baseline algorithms. The primary metric of concern is the total system cost, defined as the weighted combination of multi-user task execution latency and device energy consumption under a highly dynamic, interference-limited RIS-assisted MEC environment. The simulation settings are shown in Table. \ref{Simulation_settings}.

\begin{table}[t]
\renewcommand{\arraystretch}{1.3}
\caption{Simulation setting}
\label{Simulation_settings}
\centering
\begin{tabular}{c|c||c|c}
\hline
\bfseries Parameter & \bfseries Value & \bfseries Parameter & \bfseries Value\\
\hline
Frequency & 2.4GHz & Bandwidth & 1MHz\\
$M$ & 64 &  $K$ & 10\\
Noise & $3.16\times 10^{-11}$ & Agent type & PPO\\
Layer & 128 neurons & Activat. func. & ReLU\\
Minimum batch & 128 & Discount factor & 0.995\\
Clip factor & 0.2 & Replay memory & 10000\\
$\Delta T$ & 2s & GPU & RTX4080 \\
\hline
\end{tabular}
\end{table}

\subsubsection{Convergence speed comparison}
\begin{figure}[t]
\centering
\includegraphics[width=0.47\textwidth]{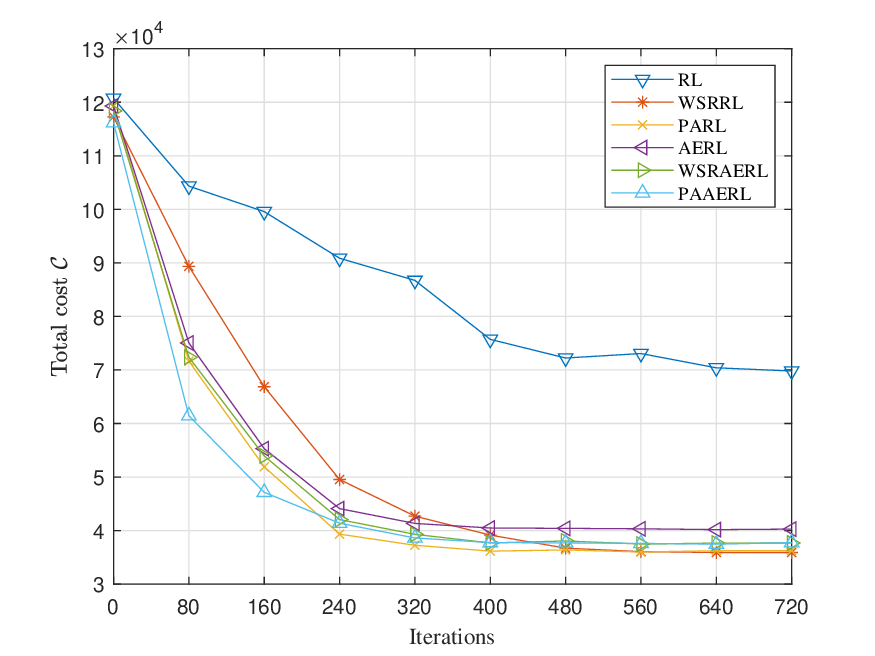}
\caption{The convergence behavior of our proposed PAAERL with benchmarks.}
\label{MEC_c_it}
\end{figure}

Fig. \ref{MEC_c_it} illustrates the total system cost as a function of the training iterations across different resource management strategies. The RL baseline exhibits the poorest performance and fails to show stable signs of convergence within the observed timeline. This behavior directly confirms the severity of the action space explosion. Without compression, the high-dimensional continuous domain of raw beamforming vectors and large-scale RIS elements forces the agent into isotropic, blind exploration, making it highly improbable to discover cooperative reward pathways. In contrast, algorithms embedded with data-driven action reduction, such as AERL, WSRAERL, and the proposed PAAERL, exhibit steep learning curves in the early training phases from 0 to 150 iterations. By utilizing the deep autoencoder to map policies onto an ultra-low-dimensional latent continuous code $\mathcal{Z}$, the agent circumvents the curse of dimensionality. The smoothed gradient updates significantly stabilize the training trajectory, allowing the policy network to rapidly isolate highly efficient macro-decisions. The steady-state costs of WSRRL and WSRAERL saturate at a significantly higher plateau compared to the Pareto-aware frameworks. This discrepancy highlights the mathematical limitation of linear scalarization in interference-limited communication links. Because the true achievable rate region under tightly coupled multi-user interference and non-convex RIS phase constraints is non-convex, the conventional weighted sum-rate formulation can only trace the outer convex hull. It systematically skips over the dented inner boundaries of the true rate region, causing critical geometric information loss and forcing the network into sub-optimal resource configurations. Both PARL and the proposed PAAERL converge to the lowest system cost boundary. This validates the effectiveness of our directional max-min scaling optimization problem $\mathcal{P}3$. By projecting weight vectors along a clean, bijective directional ray toward the Pareto-optimal frontier, our approach captures optimal operating points in both convex and non-convex rate spaces without theoretical loss of optimality.

\subsubsection{Impact of different number of users of proposed PAAERL and benchmarks}

\begin{figure}[t]
\centering
\includegraphics[width=0.47\textwidth]{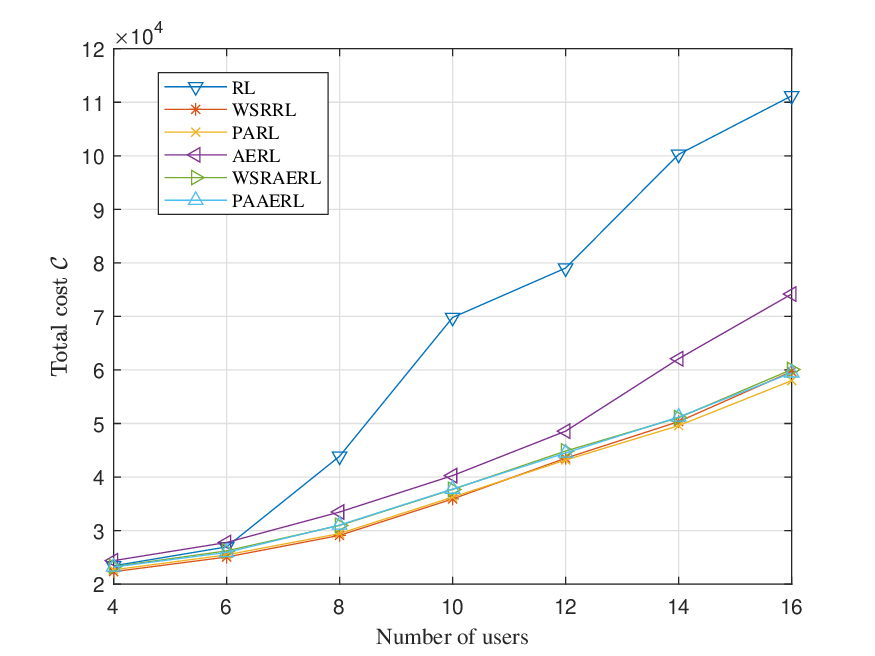}
\caption{The total cost $\mathcal C$ vs. number of users of our proposed PAAERL and benchmarks.}
\label{MEC_c_user}
\end{figure}
As observed from Fig. \ref{MEC_c_user}, the total network cost for all considered frameworks increases with the number of users $K$, which is a direct consequence of the escalated task offloading volume and aggregated multi-user interference. In particular, the conventional RL baseline exhibits catastrophic performance degradation, with its network cost accelerating dramatically as $K$ increases. This trend underlines the fatal flaw of uncompressed RL in dense networks. Because the dimensions of the active beamforming vectors $\mathbf{f}$ and the passive RIS configuration parameters are highly tied to the number of users, scaling $K$ triggers a compound explosion in the dimensionality of the action space. The conventional model-driven compression benchmarks, namely WSRRL and WSRAERL, achieve lower network costs than Conventional RL, yet their performance gap relative to the proposed PAAERL continuously widens as the network scale expands. In a sparse network with small $K$, multi-user interference is relatively mild, rendering the non-convex "dented" regions of the achievable rate region less pronounced. Thus, traditional weighted sum-rate boundary tracking remains reasonably close to optimal. However, as $K$ scales up, the mutual coupling between active transceiver beams and passive RIS reflections induces severe co-channel interference, causing the true achievable rate region to become highly non-convex. The proposed PAAERL framework consistently outperforms all other benchmark paradigms across the entire user spectrum, maintaining the lowest network cost and displaying a highly favorable, sublinear growth curve. While PARL successfully preserves optimal boundary information through the directional max-min scaling problem $\mathcal{P}3$, its execution efficiency faces bottlenecks at high values of $K$ because the policy network must still directly manipulate a $K$-dimensional priority vector $\boldsymbol{\omega}^{PT}$.

\subsubsection{Impact of different number of RIS elements of proposed PAAERL and benchmarks}

\begin{figure}[t]
\centering
\includegraphics[width=0.47\textwidth]{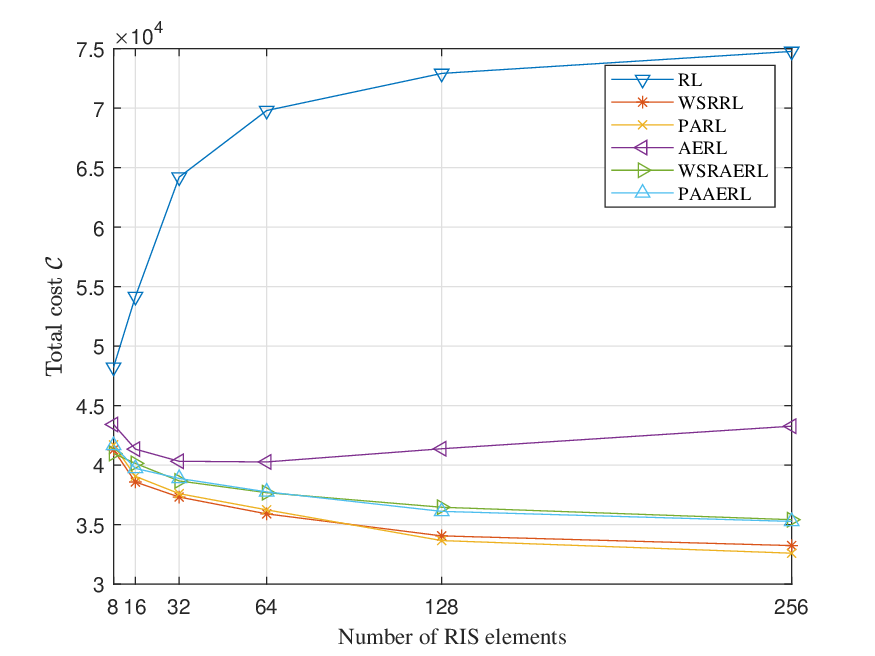}
\caption{The total cost $\mathcal C$ vs. number of RIS elements of our proposed PAAERL and benchmarks.}
\label{MEC_c_RIS}
\end{figure}

To assess the impact of passive beamforming capabilities on system-wide trade-offs, we analyze the network cost under a varying number of RIS reflecting elements $M$. Fig. \ref{MEC_c_RIS} illustrates the total system cost as a function of $M$, revealing an intriguing divergence between uncompressed model-free search paradigms and action-space-reduced frameworks. The conventional RL scheme maintains the highest network cost across the entire spectrum of $M$, followed by the AERL baseline. For conventional RL, expanding $M$ to $256$ triggers an immediate explosion in the physical layer variable domain, resulting in severe gradient dilution. The agent is forced into blind exploration, which yields highly sub-optimal beamforming and phase configurations that fail to minimize execution costs. For AERL, although the data-driven autoencoder succeeds in compressing the unconstrained action landscape into a low-dimensional code, it operates entirely without communication-theoretic or geometric optimization boundaries. Consequently, when $M$ grows large, the neural reconstruction errors within the decoder become more pronounced, causing AERL to settle into a noticeably higher cost plateau than the model-assisted algorithms. Frameworks equipped with secondary neural network compression—namely our proposed PAAERL and the hybrid WSRAERL baseline—settle into a moderate cost tier. This behavior represents a minor structural trade-off: while the deep autoencoder enables exceptionally fast policy convergence and decouples the agent from large user variables, the statistical approximation introduced during the latent space mapping $\mathcal{Z} \to \boldsymbol{\omega}^{PT}$ introduces subtle reconstruction variances. These variances prevent the agent from extracting the absolute lowest mathematical cost under highly complex channel conditions. Conversely, the un-embedded benchmarks PARL and WSRRL, which pass the priority weight vectors directly to the inner-loop convex optimization solvers without any neural bottleneck approximation, achieve the lowest total system costs.

\subsubsection{Algorithm Efficiency Comparison}

\begin{figure}[t]
\centering
\includegraphics[width=0.47\textwidth]{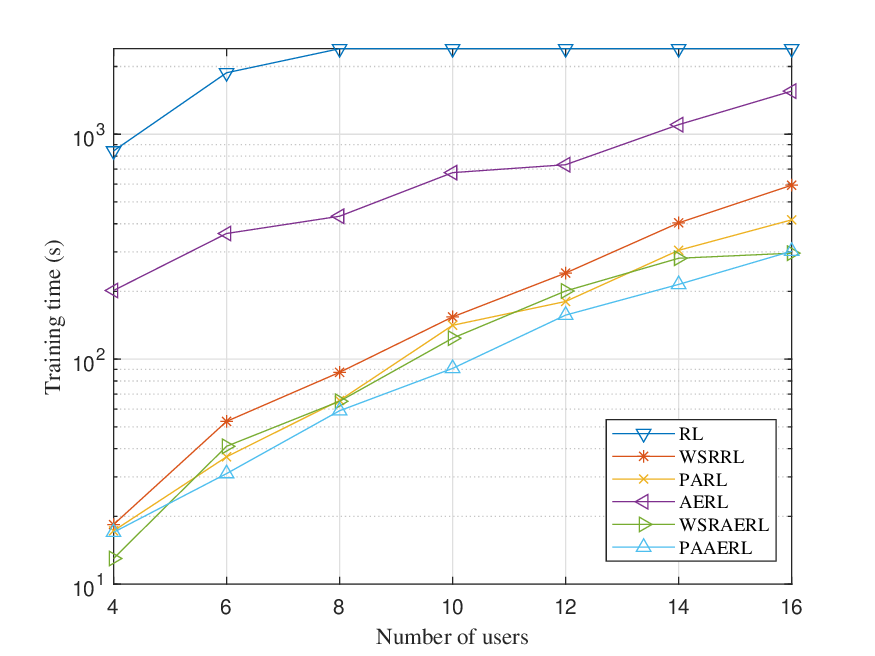}
\caption{The training time vs. number of users of our proposed PAAERL and benchmarks.}
\label{MEC_t_user}
\end{figure}

As illustrated in Fig. \ref{MEC_t_user}, the conventional RL baseline demands the highest training time across all user densities. This excessive time expenditure is a direct artifact of the action space explosion. In an uncompressed scenario, expanding the number of users requires the neural network to output exponentially larger matrix variables for multi-user active beamforming and multi-element RIS phase configurations. The baselines WSRRL, PARL, and WSRAERL yield noticeable time reductions compared to raw RL, yet they still exhibit a steep upward growth curve as $K$ scales from 4 to 16. This upward trend highlights the computational bottleneck induced by traditional iterative optimization sub-problems in the execution loop. In frameworks like WSRRL and PARL, the high-level policy outputs a $K$-dimensional weight profile. Consequently, as the number of users grows, the dimensionality of the RL action space expands proportionally. The proposed PAAERL framework consistently maintains the shortest training time across the entire user spectrum, demonstrating an exceptionally flat execution curve that scales sub-linearly with $K$. This outstanding computational efficiency is attributed to our cascaded dual-compression architecture. By cascading the data-driven autoencoder with the model-driven directional mapping, the high-level RL policy is completely isolated from the raw physical dimensions of the network. The policy network operates entirely within a steady, low-dimensional latent continuous code space $\mathcal{Z}$. This decoupling guarantees that the gradient exploration velocity remains robust even in dense user deployments.

\section{Conclusion}

%This paper presented a hierarchical RL-based algorithm that optimizes RIS systems by navigating a low-dimensional weight space. This approach maintains optimality even in non-convex scenarios and reduces the computational overhead for online IoT optimization.

This paper presented a novel hierarchical RL-based optimization framework, designated as PAAERL, to achieve real-time, online resource allocation in RIS-assisted multi-user IoT networks. To systematically break the curse of dimensionality triggered by high-dimensional active and passive beamforming configurations, we introduced a cascaded dual-compression architecture. Specifically, the framework first maps the high-dimensional physical layer parameters onto a low-scale weight space via a directional max-min scaling problem formulation. This model-driven tier mathematically guarantees strict Pareto-optimality across both convex and non-convex rate regions without any theoretical geometric information loss. To further alleviate exploration overhead in dense user deployments, a data-driven deep autoencoder was seamlessly integrated to perform a secondary action space reduction, embedding the priority profile into a highly condensed latent representation. Extensive simulations under practical multi-user MEC network conditions confirmed that our proposed PAAERL approach substantially outperforms conventional model-free RL and classic weighted sum-rate optimization baselines.

For future work, extending this dual-compression architecture to accommodate highly dynamic mobility models, such as machine-learning-driven trajectory designs for airborne RIS or UAV platforms, presents a promising and impactful research direction.

\ifCLASSOPTIONcaptionsoff
  \newpage
\fi



\bibliographystyle{IEEEtran}
\bibliography{IEEEabrv,bare_jrnl}

\end{document}